\documentstyle[psfig]{article}

\begin{document}

\def\gsim{\!\!\!\phantom{\ge}\smash{\buildrel{}\over
  {\lower2.5dd\hbox{$\buildrel{\lower2dd\hbox{$\displaystyle>$}}\over
                               \sim$}}}\,\,}
\def\kms{\rm ~km~s^{-1}}
\def\Mdot{\dot M}

\begin{center}

\Large
{\bf The Crab pulsar and its environment.}
\normalsize

\vspace{1.0cm}
Jesper Sollerman (ESO, Stockholm Observatory)
and
Veronica Flyckt (ESO,  Lule\aa~University of Technology)
\end{center}

\section{Introduction}
%--------------------%

The Crab Nebula is a supernova remnant. The supernova exploded in 
1054 AD, and was monitored by contemporary Chinese astronomers 
(see e.g., Sollerman, Kozma, \& Lundqvist 2001). 
Today the nebula offers a 
spectacular view, with a tangled web of line emitting filaments confining an 
amorphous part ghostly shining in synchrotron light. 
The beauty of the Crab makes it repeatedly appear in PR pictures, 
even from a southern observatory like the VLT (Fig. 1).

At the heart of the nebula resides the energetic 33-ms Crab pulsar. 
This m$_{V}\sim16$ 
object actually powers the whole visible nebula. 
The Crab nebula and its pulsar are among the most studied objects in the sky. 
This astrophysical laboratory still holds many secrets about
how supernovae explode and about how pulsars radiate and energize their
surrounding nebulae.

A main theme for pulsar research has been to understand the emission mechanism
for the non-thermal pulsar radiation. This is still to be accomplished. 
No comprehensive model exists that can explain all the observed features 
of the radiation. Observationally, only recently was
a broad range UV-optical spectrum of the pulsar published (Sollerman et
al. 2000). We have now extended the study into the infrared (IR).

But even if most of the research on the Crab pulsar has concerned the
radiation mechanism, almost all of the spin-down energy actually comes
out in the particle wind. This is the power source of the Crab nebula.
The stunning image of the pulsar environment 
obtained with CHANDRA (Fig. 2) captures a 
glimpse of the energetic processes at work.
Direct evidence of the pulsar activity have long been seen in the system
of moving synchrotron wisps close to the pulsar itself. 

The detailed study of the wisps was started by Scargle (1969) using
observations obtained before the authors of this article were
born. Hester et al. (1995) used the HST to study the
wisps at higher resolution, and presented their observations 
as the spectacular 'The Crab Movie'\footnote{http://oposite.stsci.edu/pubinfo/pr/1996/22.html}, where the constant activity in the
region around the pulsar is highlighted. 

The most stunning
discovery in these HST images was the knot sitting just 0.6 arcseconds
from the pulsar. At 2 kpc this amounts to a
projected distance of only 1000 AU. Hester et al. 
interpreted this feature as a shock in the pulsar polar wind.
Our IR observations also allowed us to have a look 
at these manifestations of the magnetic relativistic wind from the pulsar.

\section{IR photometry, reductions and results}
IR imaging in the short wavelength (SW) mode of ISAAC was obtained in 
service mode on VLT on October 13, 2000. 
The exposures were taken in $J$s, $H$ and
$K$s with a total exposure time of 156 seconds per band.  
The main goal of these short
exposures was to properly calibrate our IR spectroscopy. However, the
image quality provided by Paranal also allowed a detailed
view of the central region of the Crab nebula.
The near-IR images are displayed next to each other in Figure 3.

Photometry was obtained of the Crab pulsar and some of the stars in
the field using PSF-fitting (DAOPHOT). 
We estimate that our magnitude measurements of the pulsar are correct to
about 0.05 magnitudes.

In Figure 4 we plot our measurements as de-reddened fluxes together with
the optical-UV data from Sollerman et al. (2000).
We note that our IR fluxes deviate significantly from the most recent
results published by Eikenberry et al. (1997). Our measurements give a
fainter pulsar by some 0.3 magnitudes. 

The pulsar magnitudes of Eikenberry et al. 
agree with those in the  2MASS point source
catalogue. As our relatively isolated standards in the field show
good agreement with 2MASS, we do not think 
the difference in pulsar magnitudes
is due to an offset in
the zero-point. 
Instead, our measurements of the Crab pulsar agree well with previous
time-resolved photometry of the Crab (Penny 1982; Ransom et al. 1994). 
Such measurements generally use a large aperture
and simply assume any non-varying contribution to be due to the
background. By integrating under the pulsar light curve, they measure
only the pulsating contribution of the flux.  Our ISAAC photometry
has excellent signal and image quality.  In the complex region around
the pulsar, this significantly improves the background subtraction. In
particular, PSF subtraction excludes contributions from the wisps and the
nearby knot. We believe that the difference with 2MASS is
simply a matter of resolution, and that we are now able to subtract
virtually all background from the pulsar.

The 3-color image of the central parts of the Crab (Fig. 5)
is color coded with $J$s=blue, $H$=green and $K$s=red.  
In an attempt to keep some physical information in this image, 
the individual frames were scaled to
make an object with a flat de-reddened F$_{\nu}$ spectrum appear white. 

This image shows the well known features of the inner Crab. 
The wisps are clearly visible in higher detail than
previously obtained in the IR. Some filaments are also seen, most
strongly in the $J$s band. This is most likely due to the [Fe II] 
$1.26\mu$m emission line. 
The $K$s band is instead dominated by amorphous synchrotron emission (Fig. 3).
With suitable cuts the pulsar image appears
slightly elongated. This is due to the presence of the knot first
identified by Hester et al. (1995) on a HST image.  To reveal
this structure in our images we constructed and subtracted a PSF from the 
stellar images. After subtraction, the knot is clearly revealed 
in all three bands. 
A color image made out of the PSF subtracted frames is shown in Figure 6. 
The image directly gives the impression that
%It can almost be 
%seen directly on this image that
the knot is redder than for example the wisps.

Quantifying this we have estimated the magnitudes of the knot as
well as of the nearby wisp 1.  The knot
was measured within an aperture of 0.9 arcseconds and the wisp was simply
measured with an aperture of 1.2 arcseconds.  
The spectral energy distribution of the knot is shown 
in Figure 4. It is clearly red. In the $K$s-band the flux from the knot 
amounts 
to about $8\%$ of the flux of the pulsar.
The stationary wisp appears to have a flatter spectrum in this regime.

\section{Optical data from VLT and HST}

To extend the wavelength region over which to derive the spectral
characteristics of the inner Crab components, 
several optical images are available in the ESO and HST data archives.

The VLT PR image (Fig. 1) was taken with FORS2 on 10 November 1999,
just two weeks after first light. On the 
five minute $B$-band and the one minute $R$-band
exposures, we could PSF-subtract the pulsar to reveal the
knot. The knot appears clearly in both frames, and amounts to a few 
percent of the pulsar light at these wavelengths.

Data on the Crab pulsar are also available in the HST
archive.  Most of the observations are from Jeff Hester's 
comprehensive monitoring
programme of the inner parts of the nebula, which has shown just how
active the Crab nebula really is.
These frames allow a detailed study of both the spectral and temporal 
properties of the knot. In August 1995 the region was observed in 
three filters (F300W, F574M, F814W). We measured the de-reddened
spectral index for the knot to be $\alpha_{\nu}\sim-0.8$, which 
agrees well with our IR data (Fig. 4).

Furthermore, a wealth of data in the F574M filter allows a study of the 
temporal behaviour (Figure 7). 
First of all we note that the knot is indeed present 
in all frames. It thus appears quasistationary for more than six years, 
although the position appears to vary at 
the 0.1 arcsecond level. The de-reddened flux of the knot (Fig. 4) 
is measured to be $9\times10^{-28}$~ergs~s$^{-1}$~cm$^{-2}$~Hz$^{-1}$
but variations of the flux by at least $50\%$ are observed.

\section{Discussion and Implications}

For the pulsar itself we have added new information in the IR. Together with
the optical-UV data in Sollerman et al. (2000) this significantly
revises the observational basis for the pulsar emission mechanism. In fact, most
of the theoretical efforts have been based on the old
optical data from Oke (1969) and the IR continuation of Middleditch, 
Pennypacker, \& Burns (1983). Our new results 
call for a fresh look on the emission mechanism 
scenarios for young pulsars.

For the knot, we have shown that the structure is indeed quasi-stationary, 
and that the emission has a red spectrum. Few models are available for the 
knot.  Lou (1998) presented a formation scenario in terms of MHD theory, 
while Shapakidze \& Machabeli (1999) argue for a plasma mechanism. None 
of these scenarios predict a very red spectral distribution.

Another area where caution may be required is in the  
recent claims of weak and red off-pulse emission from the Crab
pulsar in the visible (Golden, Shearer, \& Beskin 1999).
It is clear that the knot close to the pulsar has to be seriously 
considered in these kinds of studies.

\section{Future plans}

The Crab pulsar and its environment continue to be the prime astrophysical 
laboratory for the study of the pulsar emission mechanism and the spin-down 
powering of pulsar nebulae.
Although much observational effort has been put into this object, 
a modern re-investigation is likely to clean up the many contradictory
measurements. Optical imaging in good seeing would require only a few minutes
with VLT, and would directly determine the knot-subtracted spectral energy 
distribution. ISAAC in the LW range can in less than 
3 hours clarify if the knot is significantly
contributing to the emission at these frequencies, and would establish if 
the IR drop of the pulsar is real.

Most interesting is the possibility to monitor the very central parts of 
the pulsar environment with NAOS/CONICA. With a resolution superseding 
HST we will be able to monitor the structures close to the pulsar, 
with 2 pixels corresponding to merely 50 AU. This would provide an unique 
opportunity to study the structure and dynamics of the inner pulsar 
wind and its interaction with the surroundings.

\section*{References}
\footnotesize

\begin{list}{}%
{\setlength {\itemindent -10mm} \setlength {\itemsep 0mm} \setlength {\parsep 0mm} \setlength {\topsep 0mm}}

\item Eikenberry, S. S., Fazio, G. G., Ransom, S. M., Middleditch, J., Kristian, J., \& Pennypacker, C. R. 1997, ApJ, 477, 465

\item Golden, A., Shearer, A., \& Beskin, G. M. 2000, ApJ, 535, 373

\item Hester, J.J, Scowen, P. A., Sankrit, R., et al. 1995, ApJ, 448, 240

\item Lou, Y.-Q. 1998, MNRAS, 294, 443

\item Middleditch, J., Pennypacker, C., \&  Burns, M. S. 1983, ApJ, 273, 261

\item Oke., J.~B. 1969, ApJ, 158, 90

\item Penny, A. J. 1982, MNRAS, 198, 773

\item Ransom, S. M., Fazio, G. G., Eikenberry, S. S.,
Middleditch, J.,  Kristian, J., Hays, K., \& Pennypacker, C. R.
1994, ApJ, 431, 43

\item Scargle, J. D. 1969, ApJ, 156, 401

\item Shapakidze, D. \& Machabeli, M. 1999, ptep.proc, 371

\item Sollerman, J., Lundqvist, P., Lindler, D., et al. 2000, ApJ, 537, 861

\item Sollerman, J., Kozma, C., \& Lundqvist, P. 2001, A\&A, 366, 197

\end{list}

\clearpage

\begin{figure}[h]
\centering
\psfig{file=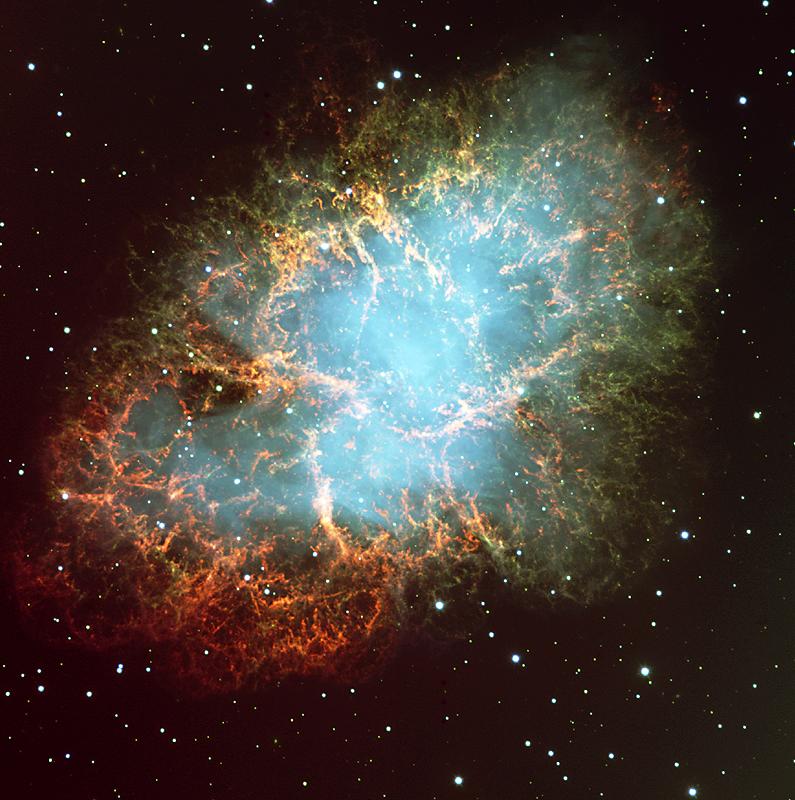,%
bbllx=10mm,bblly=105mm,bburx=215mm,bbury=200mm,width=110mm}
\vspace{2.5 cm}
\caption{ The VLT (UT2 + FORS2) view of the Crab nebula. A composite of
images in $B$(blue), $R$ and S II(red), taken in November 1999 as part of 
commissioning. ESO PR-photo 40f-99.}
\end{figure}

\begin{figure}[h]
\centering
\psfig{file=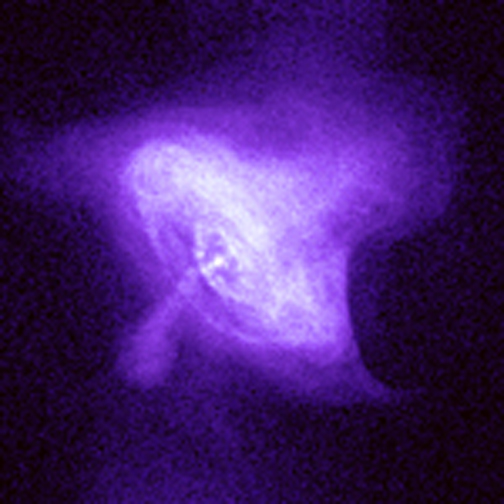,%
bbllx=10mm,bblly=105mm,bburx=215mm,bbury=200mm,width=80mm}
\vspace{4.0 cm}
\caption{The CHANDRA X-ray view of the pulsar. The spectacular torus and a 
long jet is clearly seen in this space-based 45 minute exposure from August
1999. 
The field-of-view is 2.5 arcminutes. With some imagination, the same 
structures can be seen in the VLT 5 minute $B$-band image (Fig. 1).
Photo NASA/CXC/SAO.
}
\end{figure}

\clearpage

\begin{figure}[h]
\centering
\psfig{file=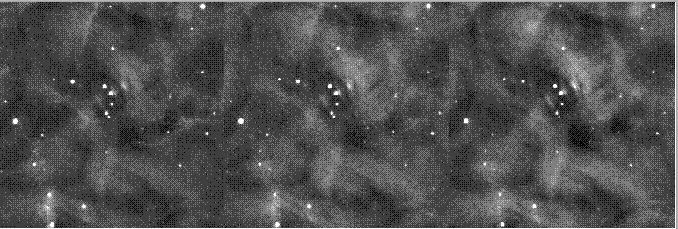,%
bbllx=0mm,bblly=0mm,bburx=540mm,bbury=180mm,width=300mm}
\vspace{5.0 cm}
\caption{The central part of the Crab nebula in the infrared, $J$s, $H$ and $K$s.
Observations with ISAAC on 13 October 2000. The pulsar is the lower right 
(South Preceding)
of the two bright objects near the center of the field. 
}
\end{figure}

\clearpage

\begin{figure}[h]
\centering
\psfig{file=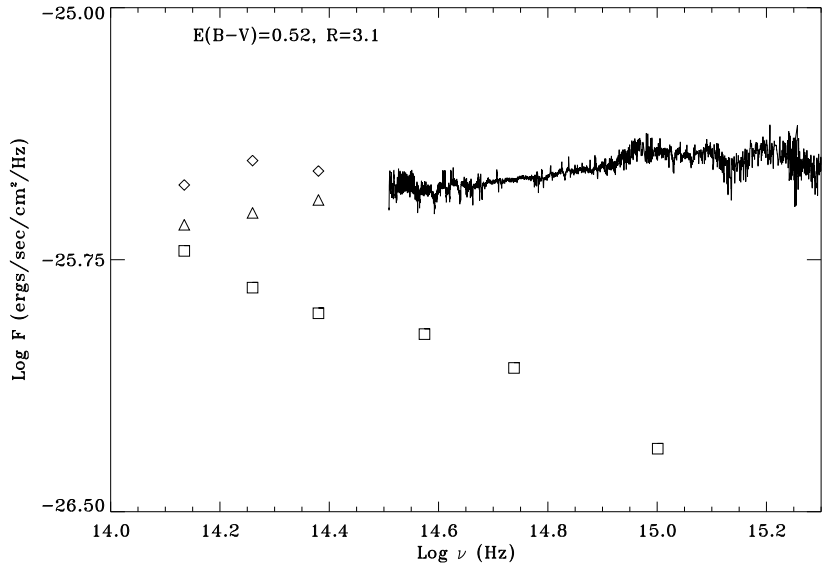,%
bbllx=10mm,bblly=105mm,bburx=215mm,bbury=300mm,width=130mm}
\vspace{3.0 cm}
\caption{Spectral energy distribution of the Crab pulsar. The optical and UV data are from Sollerman et al. (2000). The diamonds show the IR flux as published by Eikenberry et al. (1997). The triangles are our new ISAAC measurements. Also shown (squares) are the fluxes of the knot, here multiplied by a factor ten. The optical data for the knot is from HST.
All observed fluxes were de-reddened using R=3.1 and $E(B-V)$=0.52 (Sollerman et al. 2000).}
\end{figure}

\clearpage

\begin{figure}[h]
\centering
\psfig{file=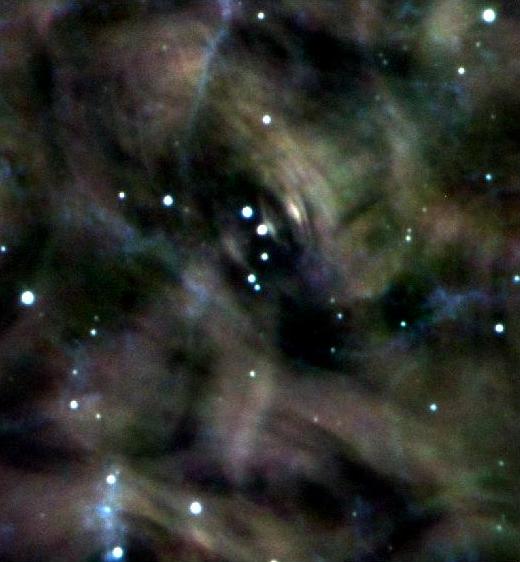,%
bbllx=10mm,bblly=105mm,bburx=215mm,bbury=200mm,width=100mm}
\vspace{4.0 cm}
\caption{The Crab nebula in the infrared. This is a color composite of the frames shown in Figure 3. North is up and East to the left.}
\end{figure}

\clearpage

\begin{figure}[h]
\centering
\psfig{file=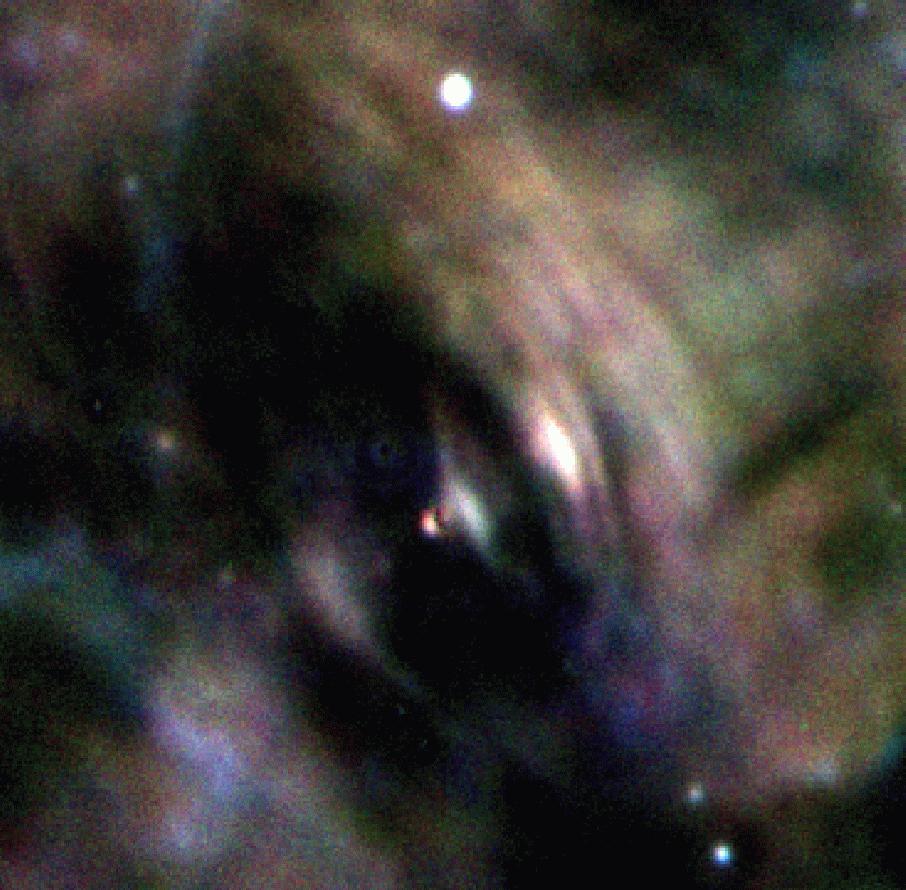,%
bbllx=10mm,bblly=105mm,bburx=215mm,bbury=200mm,width=130mm}
\vspace{3.0 cm}
\caption{The center of the Crab nebula in the infrared. After the central stars have been removed using PSF-subtraction, the knot close to the pulsar position is revealed. The FOV shown in this ISAAC image is 55.6 arcseconds, as provided by the NAOS/CONICA C12S camera.}
\end{figure}

\clearpage

\begin{figure}[h]
\centering
\psfig{file=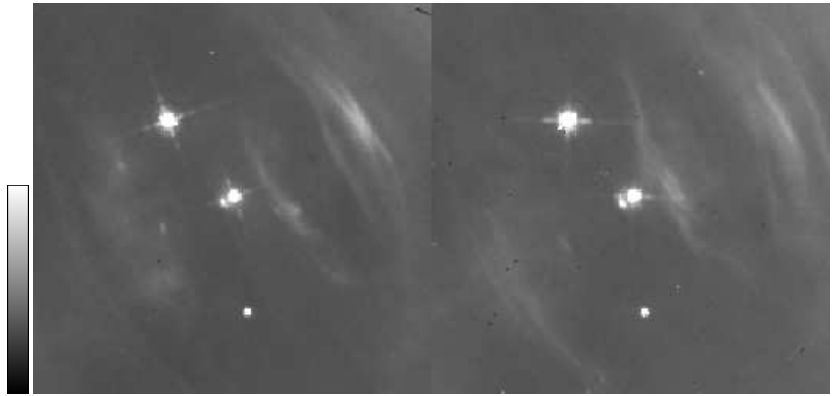,%
bbllx=0mm,bblly=0mm,bburx=540mm,bbury=270mm,width=300mm}
\vspace{1.0 cm}
\caption{HST F547M images of the very center of the nebula. The leftmost frame is obtained on  March 1994, and the right frame on October 2000, simultaneous with our IR imaging. Note that the wisp structures are very dynamic, but that the knot seen southeast of the central pulsar has persisted over more than 6 years. The FOV is 20 arcseconds, North is up and East to the left.}
\end{figure}

\end{document}